\documentclass[aps,numerical,superscriptaddress,floatfix,reprint,groupedaddres,longbibliography]{revtex4-1}
\usepackage[colorlinks=true,urlcolor={blue!75!black},citecolor={blue!75!black},linkcolor={blue!75!black}]{hyperref}
\usepackage[inline]{enumitem}
\usepackage[multidot]{grffile}
\usepackage{dcolumn}
\usepackage{bm}
\usepackage{amsmath}
\usepackage{amsfonts}
\usepackage{dsfont}
\usepackage{soul}
\usepackage{amssymb}
\usepackage{color}
\usepackage{latexsym}
\usepackage{slashed}
\usepackage{pstricks}
\usepackage{indentfirst}
\usepackage{mathrsfs}
\usepackage{tablefootnote}
\usepackage{longtable}
\usepackage{multirow}
\usepackage{epsfig,psfrag}
\usepackage{subfigure}
\usepackage{mathtools}
\usepackage{setspace}
\usepackage[utf8]{inputenc} 
\usepackage[scientific-notation=true]{siunitx} 
\usepackage{verbatim}
\graphicspath{{fig/}}
\newcommand{\uvec}{\boldsymbol}
\newcommand{\ud}{\text{d}}
\newcommand{\YC}[1]{{\color{blue} #1}}

\newcommand{\rowheight}[1]{\renewcommand{\arraystretch}{#1}}

\begin{document}
%
\title{Nucleon axial radius}
\author{Yi Chen}
\email{physchen@mail.ustc.edu.cn}
\affiliation{Department of Modern Physics, University of Science and Technology of China, Hefei, Anhui 230026, China}
\author{Yang Li}
\email{leeyoung1987@ustc.edu.cn}
\affiliation{Department of Modern Physics, University of Science and Technology of China, Hefei, Anhui 230026, China}
\author{C\'edric Lorc\'e}
\email[Corresponding author:~]{cedric.lorce@polytechnique.edu}
\affiliation{CPHT, CNRS, \'Ecole polytechnique, Institut Polytechnique de Paris, 91120 Palaiseau, France}
\author{Qun Wang}
\email{qunwang@ustc.edu.cn}
\affiliation{Department of Modern Physics, University of Science and Technology of China, Hefei, Anhui 230026, China}
\affiliation{School of Mechanics and Physics, Anhui University of Science and Technology, Huainan, Anhui 232001, China}
%
%
\begin{abstract}

We present the first systematic study of the relativistic axial-vector four-current distributions inside a nucleon. We show in particular that the slope of the axial form factor $G_A(Q^2)$ in the forward limit---conventionally denoted as $R^2_A$ in the literature---does not represent the three-dimensional mean-square axial radius in the Breit frame, but corresponds instead to a contribution to the mean-square spin radius. We derive explicit expressions for the latter in different frames and find in general additional contributions that depend on both the nucleon mass and the forward values of the axial-vector form factors $G_A(0)$ and $G_P(0)$. This provides an additional key motivation for ongoing lattice quantum chromodynamics (QCD) calculations and future experimental measurements of the induced pseudoscalar form factor $G_P(Q^2)$.

\end{abstract}
\maketitle
%
%
\textit{Introduction.} A fundamental aspect of the nucleon spin structure is captured by the axial-vector form factors (FFs), which are Lorentz-invariant quantities playing a crucial role in the weak sector, especially in neutrino scattering~\cite{Mann:1973pr,Baker:1981su,Horstkotte:1981ne,Ahrens:1986xe,MiniBooNE:2010bsu,MiniBooNE:2010xqw,MINERvA:2013bcy,MiniBooNE:2013dds,Meyer:2016oeg,Sufian:2018qtw,MINERvA:2023avz} and oscillation~\cite{NOvA:2007rmc,T2K:2011qtm,DUNE:2020ypp,Hyper-Kamiokande:2018ofw,JUNO:2015zny} experiments. Numerous theoretical investigations of these FFs based on chiral perturbation theory and phenomenological models can be found in Refs.~\cite{Bernard:1994wn,Schindler:2006it,Eichmann:2011pv,Dahiya:2014jfa,Anikin:2016teg,Mamedov:2016ype,Mondal:2019jdg,Chen:2021guo,Cheng:2022jxe,Liu:2022ekr,Borah:2024hvo,Ramalho:2024tdi}. Tremendous progress from the lattice quantum chromodynamics (QCD) side has also been reported in the last few years~\cite{Green:2017keo,Shintani:2018ozy,Jang:2019vkm,RQCD:2019jai,Park:2021ypf,Alexandrou:2021wzv,Djukanovic:2022wru,Jang:2023zts,Alexandrou:2023qbg}. For recent reviews on axial-vector FFs, see e.g.~Refs.~\cite{Bernard:2001rs,Gorringe:2002xx,Meyer:2022mix,SajjadAthar:2022pjt,Gupta:2024qip}.

Following the early works~\cite{Meissner:1986xg,Meissner:1986js}, the mean-square axial radius of the nucleon is traditionally defined in the literature as~\cite{Bernard:1992ys,A1:1999kwj,Hill:2017wgb,MINERvA:2023avz,Petti:2023abz,Kaiser:2024vbc}
\begin{equation}\label{def-Sachs-radius-Axial}
	R_A^2 \equiv -\frac{6}{G_A(0)} \frac{\ud G_A(Q^2)}{\ud Q^2}\bigg{|}_{Q^2=0},
\end{equation}
where $G_A(Q^2)$ is the nucleon axial FF. This definition is based on an analogy with the standard (conventional) expression for the mean-square charge radius~\cite{Ernst:1960zza,Sachs:1962zzc,Gao:2021sml}
\begin{equation}\label{def-Sachs-radius-Electric}
	\langle r_E^2 \rangle =-\frac{6}{G_E(0)} \frac{\ud G_E(Q^2)}{\ud Q^2}\bigg{|}_{Q^2=0},
\end{equation}
where $G_E(Q^2)$ is the nucleon Sachs electric FF.

It should be stressed that Eq.~\eqref{def-Sachs-radius-Electric} is not merely a definition, but the result of evaluating the $r^2$-moment of the conventional electric charge distribution of the nucleon in the Breit frame (BF)~\cite{Sachs:1962zzc}. In this particular case, it turns out that the conventional mean-square charge radius can be simply expressed in terms of the slope of the electric FF at $Q^2=0$. However, the relations between genuine mean-square radii and FFs are in general more complicated. For example, the mean-square mass radius of a spin-$0$ target (of mass $M$) is given by~\cite{Polyakov:2018zvc}
\begin{equation}\label{massrad}
    \langle r_\text{mass}^2 \rangle =-6\,\frac{\ud A(Q^2)}{\ud Q^2}\bigg{|}_{Q^2=0}-\frac{3D(0)}{2M^2}-\frac{3}{4M^2},
\end{equation}
where $A(Q^2)$ and $D(Q^2)$ are gravitational FFs that parametrize the energy-momentum tensor of the system, and the last term is a relativistic recoil contribution that was neglected in Ref.~\cite{Polyakov:2018zvc}. Similar expressions have been found for spin-$\frac{1}{2}$~\cite{Lorce:2018egm}, spin-$1$~\cite{Cosyn:2019aio} and spin-$\frac{3}{2}$~\cite{Kim:2020lrs} targets. In the case of the mechanical radius, the expression does not even involve any FF derivative~\cite{Polyakov:2018zvc}.

In view of these observations and the prevailing misconceptions in the literature, the aim of the present Letter is to investigate in detail whether the widespread interpretation of $R_A^2$ given by Eq.~(\ref{def-Sachs-radius-Axial}) as the nucleon mean-square axial radius is justified or not. In particular, we demonstrate that the nucleon mean-square spin radius, defined in terms of both the axial and induced pseudoscalar FFs, is a natural and physically meaningful 3D mean-square radius that better characterizes the size of the weak content of the nucleon.
\newline

%
\textit{Mean-square radii.} Fundamentally, the notion of mean-square radius characterizes the extension of a spatial distribution $\rho(\uvec r)$, and is defined as~\cite{Chen:2023dxp}
    \begin{equation}\label{def-RMS}
	\langle r^2 \rangle \equiv \frac{\int \ud^3 r\,r^2 \rho(\uvec r) }{\int \ud^3 r\,\rho(\uvec r)}=-\frac{1}{\tilde\rho(\uvec 0)} \uvec\nabla^2_{\uvec\Delta} \tilde\rho(\uvec\Delta)\Big|_{\uvec\Delta=\uvec 0},
\end{equation}
where
\begin{equation}
    \tilde\rho(\uvec\Delta)=\int\ud^3r\,e^{i\uvec\Delta\cdot\uvec r}\,\rho(\uvec r).
\end{equation}
This inverse Fourier transform is usually identified with some BF amplitude, i.e.~a matrix element where the initial and final momenta satisfy $\uvec p_B=-\uvec p'_B=-\uvec\Delta/2$. 

For example, the conventional electric charge distribution of the nucleon~\cite{Sachs:1962zzc} is defined as the Fourier transform of the following BF amplitude
\begin{equation}
    \tilde J^0_B(\uvec\Delta)\equiv\frac{\langle p'_B,\uparrow|j^0(0)|p_B,\uparrow\rangle}{2M}=e\,G_E(\uvec\Delta^2),
\end{equation}
where $j^\mu(x)$ is the electromagnetic four-current operator, $|p,s\rangle$ is a covariantly normalized momentum eigenstate with canonical polarization $s$, and $e$ is the positive unit of electric charge. Setting $\tilde\rho=\tilde J^0_B$ in Eq.~\eqref{def-RMS} leads directly to Eq.~\eqref{def-Sachs-radius-Electric}, since in the BF we can write $Q^2=\uvec\Delta^2$.

This is to be contrasted with the phase-space formalism~\cite{Lorce:2020onh,Chen:2022smg,Chen:2023dxp}, where the intrinsic relativistic electric charge distribution of the nucleon is unambiguously defined as the Fourier transform of
\begin{equation}\label{Vector_current}
    \tilde{\mathcal J}^0_B(\uvec\Delta)\equiv\frac{\langle p'_B,\uparrow|j^0(0)|p_B,\uparrow\rangle}{2P^0_B}=e\,\frac{M}{P^0_B}\,G_E(\uvec\Delta^2),
\end{equation}
in agreement with older analyses~\cite{Yennie:1957rmp,Friar:1975pp} and the requirement that the total electric charge must transform as a Lorentz scalar~\cite{Lorce:2020onh}. The BF energy being given by $P^0_B=p'^0_B=p^0_B=M\sqrt{1+\tau}$ with $\tau\equiv Q^2/(4M^2)$, one concludes that the relativistic mean-square charge radius of the proton is in fact given by~\cite{Chen:2023dxp,Friar:1997js}
\begin{equation}
    \langle r_\text{ch}^2 \rangle=\langle r_E^2 \rangle+\frac{3}{4M^2}.
\end{equation}
The additional term, known as the Darwin-Foldy term~\cite{Yennie:1957rmp,Foldy:1950ws,Foldy:1952df,Friar:1997js}, see also the Eqs.~(5b) and (5c) of Ref.~\cite{Friar:1997js}, actually appeared in Ref.~\cite{Ernst:1960zza} but was later discarded by hand by the authors to comply with the conventions in force in atomic spectroscopy~\cite{Miller:2018ybm}.
\newline

%
\textit{3D Breit frame distributions.} Using solely fundamental symmetries (namely Poincar\'e symmetry including parity, time-reversal and charge-conjugation symmetries, and the $G$-parity invariance of QCD), the matrix elements of the axial-vector four-current operator $j_5^\mu(x)=\overline\psi(x)\gamma^\mu \gamma_5 \psi(x)$ for a given quark flavor and for a generic spin-$\frac{1}{2}$ target can be parametrized as~\cite{Bernard:2001rs,Jang:2019vkm}
\begin{equation}
	\begin{aligned}\label{MatrElem-AxialVect}
		&\langle p',s'|j^\mu_5(0)|p,s\rangle= \\
  &\quad \overline u(p',s')\left[\gamma^\mu\gamma_5\, G_A(Q^2) + \frac{\Delta^{\mu}\gamma_5}{2 M}\,G_P(Q^2) \right]u(p,s),
	\end{aligned}
\end{equation}
where $G_P(Q^2)$ is the induced pseudoscalar FF, and $\Delta^\mu=p'^\mu-p^\mu$ is the four-momentum transfer.

Similarly to Eq.~\eqref{Vector_current}, we define the axial-vector BF amplitudes as
\begin{equation}
    \tilde {\mathcal A}^\mu_B(\uvec\Delta)\equiv\frac{\langle p'_B,\uparrow|j_5^\mu(0)|p_B,\uparrow\rangle}{2P^0_B}
\end{equation}
with the nucleon spin polarized along an arbitrary direction denoted by the unit vector $\hat {\uvec s}$. We obtain
\begin{equation}
	\begin{aligned}\label{3DBF-ampl}
		\tilde{\mathcal A}^0_B &= 0,\\
		\tilde{\mathcal A}^i_B
		&=\left[ \hat s^i - \frac{\Delta^i(\uvec\Delta \cdot \hat{\uvec s}) }{4P_B^0 (P_B^0+M)} \right]G_A(\uvec\Delta^2)\\
		&\qquad - \frac{\Delta^i(\uvec\Delta \cdot \hat{\uvec s}) }{4MP^0_B}\,G_P(\uvec\Delta^2).
	\end{aligned}
\end{equation}
Contrary to the electromagnetic case, we see that the axial charge density vanishes identically in the BF, hence there is no nontrivial axial charge radius. This can be understood from the fact that the electromagnetic four-current is timelike, whereas the axial-vector four-current is spacelike.

Although the axial charge density vanishes identically in the BF, it is still possible to provide a density interpretation for the spatial part of the axial-vector current. Using the QCD equations of motion, the rank-$3$ canonical spin tensor operator can be expressed as $S^{\mu\alpha\beta}=\tfrac{1}{2}\epsilon^{\mu\alpha\beta\lambda} j_{5\lambda}$ with $\epsilon_{0123}=+1$~\cite{Leader:2013jra}. The spin density operator is then given by $S^i=\frac{1}{2}\epsilon^{ijk}S^{0jk}=\tfrac{1}{2} j_5^i$. As a result, we can define a meaningful 3D nucleon spin distribution in the BF as follows~\cite{Lorce:2017wkb}
\begin{equation}\label{3D-spin-dist}
    S^i_B(\uvec r)=\frac{1}{2}\int\frac{\ud^3\Delta}{(2\pi)^3}\,e^{-i\uvec\Delta\cdot\uvec r}\,\tilde{\mathcal A}^i_B(\uvec\Delta).
\end{equation}
The 3D nucleon spin radius $r_S$ can then be unambiguously defined as $r_S=\sqrt{\langle r_\text{spin}^2 \rangle}$, with
\begin{equation}
	\begin{aligned}\label{def-new-spinMSR}
		\langle r_\text{spin}^2 \rangle
		&\equiv \frac{\int \ud^3 r\,r^2  \, \hat{\uvec s} \cdot\uvec S_B(\uvec r) }{\int \ud^3 r\, \hat{\uvec s} \cdot\uvec S_B (\uvec r)}\\
		&=R_A^2  + \frac{1}{4M^2 }\left(1+\frac{2\,G_P(0)}{G_A(0)}\right).
	\end{aligned}
\end{equation}
Besides the $R^2_A$ term coming from the slope of the axial FF at $Q^2=0$ given by Eq.~\eqref{def-Sachs-radius-Axial}, we find two additional contributions associated with the $\Delta^i(\uvec\Delta\cdot \hat{\uvec s} )$ terms in Eq.~\eqref{3DBF-ampl}. The latter can be understood as relativistic effects~\cite{Lorce:2021gxs,Chen:2023dxp} which can formally be ignored only if we consider the static limit $M\to\infty$.

\begin{figure}[t!]
    \centering
    {\includegraphics[angle=0,scale=0.452]{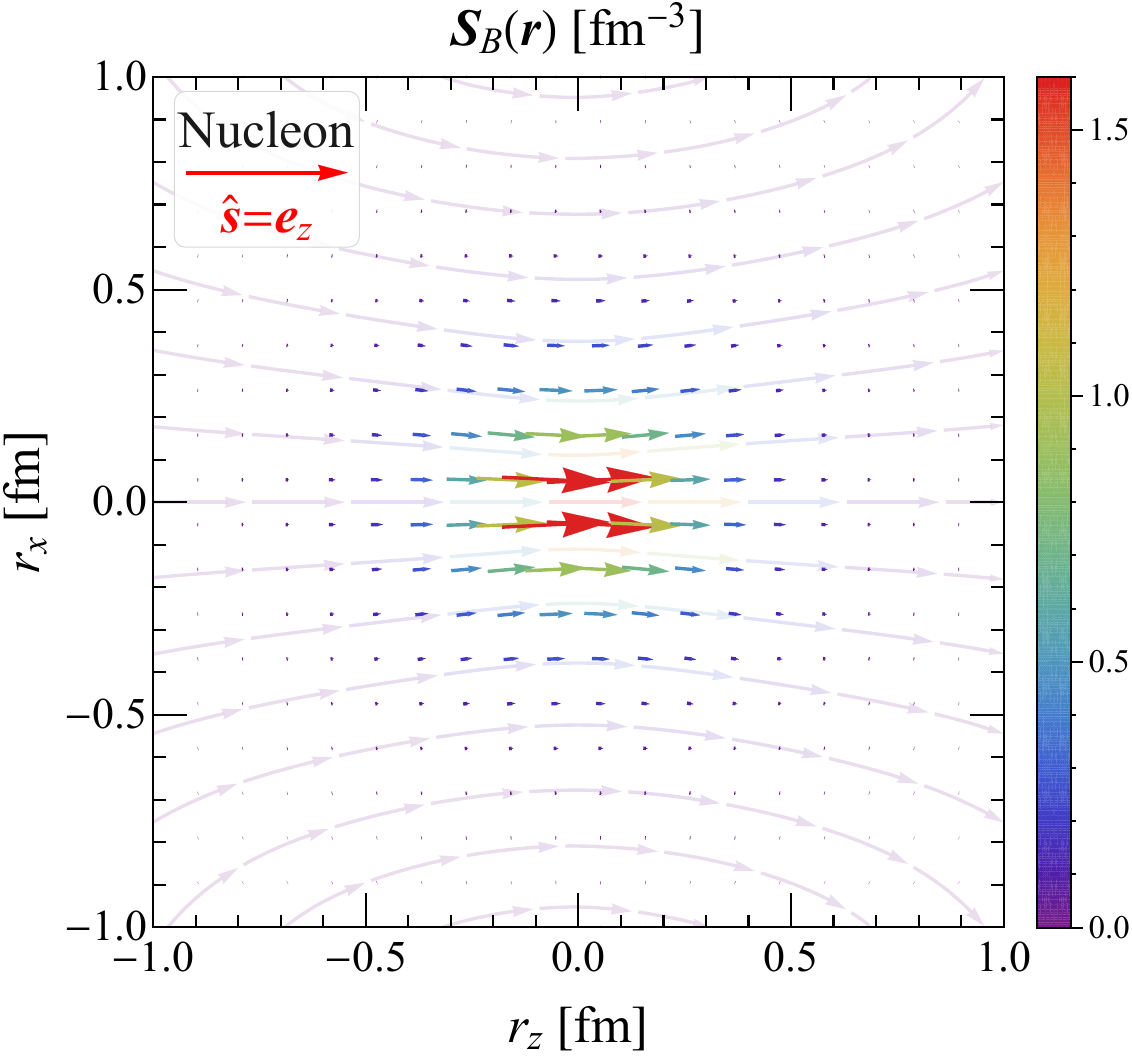}}
    \caption{The 3D relativistic spin distribution $\uvec S_B(\uvec r)$ from Eq.~\eqref{3D-spin-dist} in the $r_y=0$ plane for a longitudinally polarized nucleon, using the nucleon axial-vector FFs~\eqref{FFparamdip} and parameters from Table~\ref{Tab_FFs_Parameters}.}
    \label{Fig_3DBFSvT}
    \vspace{-3mm}
\end{figure}

In Fig.~\ref{Fig_3DBFSvT}, we illustrate the 3D spin distribution~\eqref{3D-spin-dist} inside the nucleon in the $r_y=0$ plane using the nucleon axial-vector FFs $G_A(Q^2)\equiv \sum_f G_A^f(Q^2)$ and $G_P(Q^2)\equiv \sum_f G_P^f(Q^2) $ from recent lattice QCD calculations~\cite{Alexandrou:2021wzv}, where $G_X^f(Q^2)$ for $X=A,P$ and $f=u,d$ are parametrized as
\begin{equation}\label{FFparamdip}
		G_X^f(Q^2)= \frac{G_X^f(0)}{\left(1+ \frac{Q^2}{ (m_X^f)^2 } \right)^2 }
\end{equation}
with the dipolar parameters given in Table~\ref{Tab_FFs_Parameters}. It should be noted that the contributions from heavier quarks (e.g., $s$ and $c$ quarks) are not included in our numerical calculations due to their small contributions but large uncertainties~\cite{Alexandrou:2021wzv}. We find in particular that $r_S \approx 0.916~\text{fm}$ and $R_A \approx 0.567~\text{fm}$ by using the lattice parametrization of $G_X^f(Q^2)$ from Ref.~\cite{Alexandrou:2021wzv}. This indicates that the relativistic contribution $\frac{1}{4M^2 }\left(1+\frac{2\,G_P(0)}{G_A(0)}\right)$ in Eq.~\eqref{def-new-spinMSR} is not small.
\newline

\begin{table}[t!]
	\rowheight{1.2}
	\caption[]{The dipole model parametrization of the axial and induced pseudoscalar form factors of the $u$ and $d$ quarks for the nucleon from Ref.~\cite{Alexandrou:2021wzv}.}
	\begin{center}
		\vspace{-2mm}
		\begin{ruledtabular}
			\begin{tabular}{ c c c c }
				$G_A^u(0)\qquad$ & $G_A^d(0)\qquad$ & $m_A^u~\text{[GeV]}\qquad$ & $m_A^d~\text{[GeV]}$\\
				\hline
				$0.859\qquad$ & $-0.423\qquad$ & $1.187\qquad$ & $1.168$ \\
				\hline
				\hline
				$G_P^{u}(0)\qquad$ & $G_P^d(0)\qquad$ & $m_P^u~\text{[GeV]}\qquad$ & $m_P^d~\text{[GeV]}$ \\
				\hline
				$125\qquad$ & $-115\qquad$ & $0.193\qquad$ & $0.191$ 
			\end{tabular}
		\end{ruledtabular}
		\vspace{-5mm}
		\label{Tab_FFs_Parameters}
	\end{center}
\end{table}

%
\textit{2D light-front distributions.} Although BF distributions provide us the best proxy for picturing in 3D a system sitting in average at rest around the origin, their lack of strict probabilistic interpretation is often regarded as a deficiency~\cite{Miller:2010nz}. By focusing on a Galilean subgroup of the Lorentz group, one can define alternative densities with strict probabilistic interpretation~\cite{Burkardt:2002hr} by using the light-front (LF) formalism, see e.g. Refs.~\cite{Miller:2010nz,Miller:2007uy,Carlson:2007xd,Freese:2023jcp,Freese:2023abr}.

In the LF formalism~\cite{Brodsky:1997de}, a four-vector is represented by the set of components $a^\mu=[a^+,a^-,\uvec a_\perp]$ with $a^\pm\equiv(a^0\pm a^3)/\sqrt{2}$. Since the scalar product takes the form $p\cdot x=p^-x^++p^+x^--\uvec p_\perp\cdot\uvec x_\perp$, treating $x^+$ as the LF time variable implies that $p^-$ corresponds to the LF energy, given by $(\uvec p^2_\perp+M^2)/(2p^+)$ according to the on-mass-shell relation $p^2=M^2$.

Similarly to the electromagnetic distributions studied in Refs.~\cite{Chen:2023dxp,Lorce:2020onh,Chen:2022smg}, the axial-vector LF distributions are defined via the following two-dimensional Fourier transform
\begin{equation}\label{LF-def}
		\mathcal A^\mu_\text{LF}(\uvec b_\perp;P^+)=\int\frac{\ud^2\Delta_\perp}{(2\pi)^2}\,e^{-i\uvec\Delta_\perp\cdot\uvec b_\perp}\,\tilde{\mathcal A}_\text{LF}^\mu(\uvec\Delta_\perp;P^+),
\end{equation}
where the amplitudes with given initial and final LF helicities
\begin{equation}
   \tilde{\mathcal A}_\text{LF}^\mu(\uvec\Delta_\perp;P^+)\equiv\frac{ {_\text{LF}}\langle p',\lambda'|j_5^\mu(0)|p,\lambda\rangle_\text{LF} }{2P^+}
\end{equation}
are evaluated in the so-called symmetric LF frame specified by $\uvec P_\perp=(\uvec p'_\perp+\uvec p_\perp)/2=\uvec 0_\perp$ and $\Delta^+=0$. Note that the latter condition amounts to an integration over the longitudinal LF coordinate $r^-$ in position space.

Using the LF Dirac spinors in the parametrization~\eqref{MatrElem-AxialVect}, we find
\begin{equation}
	\begin{aligned}\label{2DLF-ampl}
		\tilde{\mathcal A}_\text{LF}^+ 
		&= (\sigma_z)_{\lambda'\lambda}\,G_A(\uvec\Delta_\perp^2),\\
		\tilde{\mathcal A}_\text{LF}^- 
		&= -\frac{P^-}{P^+}\, (\sigma_z)_{\lambda'\lambda}\, G_A(\uvec\Delta_\perp^2),\\
		\tilde{\mathcal A}^i_{\perp,\text{LF}}
		&= \frac{M}{P^+}\left[(\sigma^i_\perp )_{\lambda'\lambda}+ \frac{i\tilde \Delta^i_\perp}{2M}\, \delta_{\lambda'\lambda} \right]G_A(\uvec\Delta_\perp^2)\\
		&\quad -\frac{\Delta^i_\perp\left(\uvec\Delta_\perp \cdot \uvec\sigma_\perp \right)_{\lambda'\lambda}}{4MP^+}\,G_P(\uvec\Delta_\perp^2),
	\end{aligned}
\end{equation}
where $\sigma^i$ are the Pauli matrices and $\tilde{\uvec \Delta}_\perp \equiv \uvec e_z \times \uvec\Delta_\perp$. The LF axial charge distribution $\mathcal A^+_\text{LF}(\uvec b_\perp)$ turns out to be $P^+$-independent. It does not vanish in the case of a longitudinally polarized nucleon and is interpreted in the LF formalism as the quark helicity distribution~\cite{Burkardt:2002hr,Diehl:2005jf}.

The 2D transverse mean-square helicity (or axial) radius is naturally defined on the LF for $\lambda'=\lambda$ as~\cite{Diehl:2005jf}
\begin{equation}\label{2DLF-ACR}
		\langle b_A^2 \rangle_\text{LF}
		= \frac{\int\ud^2b_\perp\, b^2_\perp \,\mathcal A^+_\text{LF}(\uvec b_\perp)}{\int\ud^2b_\perp\,\mathcal A^+_\text{LF}(\uvec b_\perp)} = \frac{2}{3}\, R_A^2.
\end{equation}
Since LF distributions are two-dimensional, it is tempting to interpret $R^2_A$ as the corresponding 3D mean-square axial radius, e.g.~based on the Abel-Radon transformation~\cite{Panteleeva:2021iip,Kim:2021jjf}. This would be justified if we could demonstrate that the LF axial charge distribution is the projection of a spherically symmetric 3D distribution onto the transverse plane. Unfortunately, 3D LF distributions with strict probabilistic interpretation do not exist because of the necessary $\Delta^+=0$ constraint~\cite{Freese:2021mzg}.

\begin{figure}[t!]
    \centering
    {\includegraphics[angle=0,scale=0.452]{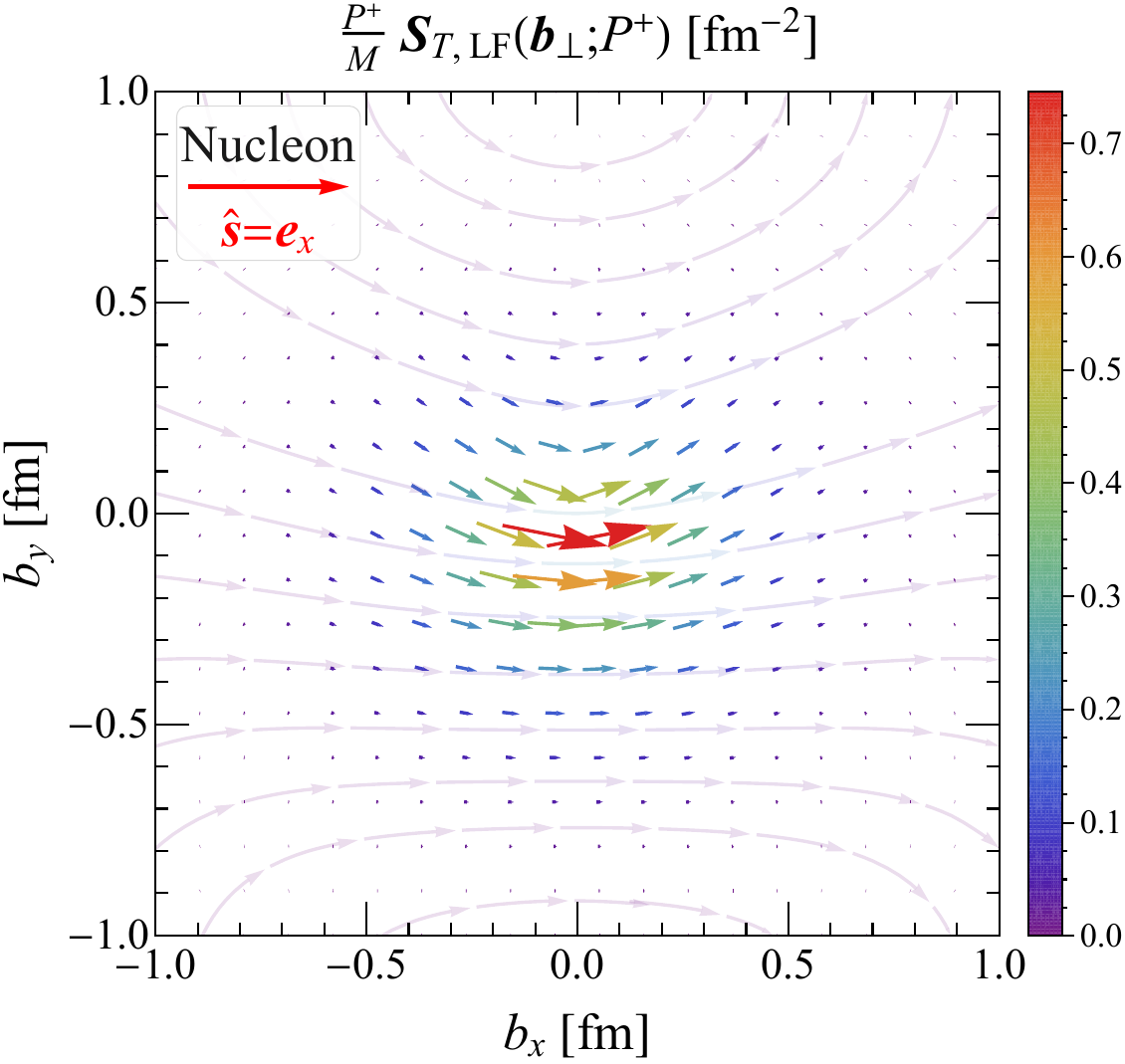}}
    \caption{The scaled 2D transverse LF spin distribution $\frac{P^+}{M} \uvec S_{T,\text{LF}}(\uvec b_\perp;P^+ )=\frac{P^+}{2M} \uvec{\mathcal A}_{\perp,\text{LF}}(\uvec b_\perp;P^+ )$ from Eq.~\eqref{LF-def} for a transversely polarized nucleon, using the nucleon axial-vector FFs~\eqref{FFparamdip} and parameters from Table~\ref{Tab_FFs_Parameters}.}
    \label{Fig_2DLFSvT}
    \vspace{-3mm}
\end{figure}

Alternatively, one can consider the longitudinal LF spin distribution given by $S_{L,\text{LF}}\equiv\frac{1}{4}\text{Tr}[\sigma_z\, \mathcal A^+_\text{LF}]$, which is nothing other than half of the light-front helicity distribution. Since by definition the (transverse) mean-square radius does not depend on the overall normalization of the distribution, we obtain naturally
\begin{equation}\label{2DLF-SpinMSR-L}
    \langle b_{\text{spin},L}^2 \rangle_\text{LF}=\langle b_A^2 \rangle_\text{LF}=\frac{2}{3}\, R_A^2.
\end{equation}
If we use instead the transverse LF spin distribution $S_{T,\text{LF}}\equiv\frac{1}{4}\text{Tr}[\uvec\sigma_\perp\cdot\uvec{\mathcal A}_{\perp,\text{LF}}]=\hat{\uvec s}_\perp\cdot\uvec S_{T,\text{LF}}$, where the numerical result of $\uvec S_{T,\text{LF}}(\uvec b_\perp;P^+)$ is illustrated in Fig.~\ref{Fig_2DLFSvT}, we find
\begin{equation}\label{2DLF-SpinMSR-T}
        \langle b_{\text{spin},T}^2 \rangle_\text{LF}= \frac{2}{3}\,R_A^2 + \frac{1}{2 M^2}\frac{ G_P(0)}{G_A(0)}.
\end{equation}
Note that the $P^+$-dependence of $S_{T,\text{LF}}$ enters merely as a global factor and drops out when calculating the transverse mean-square radius. The difference between longitudinal and transverse LF (transverse) mean-square spin radii is of relativistic origin (\ref{2DLF-SpinMSR-T}), and disappears in the static limit $M\to\infty$.
\newline

%
\textit{2D elastic frame distributions.} BF and LF axial-vector distributions at first appearance look somewhat different. To understand how they are related to each other, we can use the phase-space formalism and define 2D elastic frame (EF) distributions~\cite{Lorce:2017wkb,Lorce:2018egm}. Besides transverse position, these time-independent distributions depend also on the average momentum $\uvec P=(\uvec p'+\uvec p)/2\equiv P_z \uvec e_z$ of the target, which defines the $z$-axis without loss of generality~\cite{Lorce:2020onh}. Taking the limit $P_z\to 0$ we obtain naturally the projection of the BF distributions onto the transverse plane, while taking the infinite momentum limit $P_z\to\infty$ we recover basically the LF distributions~\cite{Lorce:2020onh,Chen:2022smg,Chen:2023dxp}.

Similarly to the axial-vector LF distributions~\eqref{LF-def}, the axial-vector EF distributions are defined via the following two-dimensional Fourier transform
\begin{equation}\label{EF-def}
		\mathcal A^\mu_\text{EF}(\uvec b_\perp;P_z)=\int\frac{\ud^2\Delta_\perp}{(2\pi)^2}\,e^{-i\uvec\Delta_\perp\cdot\uvec b_\perp}\,\tilde{\mathcal A}_\text{EF}^\mu(\uvec\Delta_\perp;P_z),
\end{equation}
where the EF amplitudes in polarization space
\begin{equation}
   \tilde{\mathcal A}_\text{EF}^\mu(\uvec\Delta_\perp;P_z)\equiv\frac{\langle p',s'|j_5^\mu(0)|p,s\rangle}{2P^0}
\end{equation}
are evaluated in a generic frame where $\uvec P_\perp=\uvec 0_\perp$ and $\Delta_z=0$ so as to ensure that no energy is transferred to the system. We can therefore write $p'^0=p^0=P^0=\sqrt{M^2(1+\tau)+P_z^2}$. 

Evaluating the matrix elements~\eqref{MatrElem-AxialVect} in a generic EF, we find
\begin{equation}
	\begin{aligned}\label{2DEF-ampl}
		\tilde{\mathcal A}_\text{EF}^0  
		&= \frac{P_z}{P^0}\, (\sigma_z)_{s's}\, G_A(\uvec\Delta_\perp^2) ,\\
		\tilde{\mathcal A}_{z,\text{EF}}   
		&= (\sigma_z)_{s's}\, G_A(\uvec\Delta_\perp^2),\\
		\tilde{\mathcal A}^i_{\perp,\text{EF}} 
		&=\\
  \frac{\sqrt{P^2}}{P^0}&\left[(\sigma^i_\perp)_{s's}\,\cos\theta-\frac{i\tilde \Delta^i_\perp}{2M\sqrt{\tau}}\,\delta_{s's}\,\sin\theta\right]G_A(\uvec\Delta^2_\perp)\\
  &- \frac{\Delta^i_\perp \left( \uvec\Delta_\perp \cdot \uvec\sigma_\perp \right)_{s's} }{4P^0}\,\left[\frac{G_A(\uvec\Delta_\perp^2)}{P^0+M}+\frac{G_P(\uvec\Delta_\perp^2)}{M}\right],
	\end{aligned}
\end{equation}
with the Wigner rotation given by~\cite{Chen:2023dxp,Lorce:2022jyi,Chen:2022smg}
\begin{equation}\label{WignerAngleSinCos}
	\begin{aligned}
		\cos\theta&=\frac{P^0+M(1+\tau)}{(P^0+M)\sqrt{1+\tau}},\\ 
        \sin\theta&=-\frac{\sqrt{\tau}\,P_z}{(P^0+M)\sqrt{1+\tau}}.
	\end{aligned}
\end{equation}
Above results nicely agree with the expected Lorentz transformation law for amplitudes from the BF to a generic EF~\cite{Jacob:1959at,Durand:1962zza}
\begin{equation}
	\begin{aligned}\label{CovaLorentzTrans-Spinj}
		\langle p',s'|j_5^\mu(0)|p,s\rangle 
		&= \sum_{s_{B}',s_{B}} D^{\dag(1/2)}_{s's_{B}'}(p_{B}',\Lambda)D^{(1/2)}_{s_{B}s}(p_{B},\Lambda)\\ &\quad \times \Lambda^{\mu}_{\phantom{\mu}\nu} \,\langle p_{B}',s_{B}'|j_5^{\nu}(0)|p_{B},s_{B}\rangle,
	\end{aligned}
\end{equation}
where $D^{(1/2)}$ is the Wigner spin rotation matrix for a spin-$\frac{1}{2}$ system~\cite{Chen:2023dxp,Chen:2022smg}. It is straightforward to see that the EF amplitudes~\eqref{2DEF-ampl} reduce to the BF ones~\eqref{3DBF-ampl} with $\Delta_z=0$ when $P_z\to 0$, and to the LF ones~\eqref{2DLF-ampl} up to some convention-dependent factors when $P_z\to\infty$~\cite{Chen:2022smg,Chen:2023dxp}. 

When $P_z>0$ and the target is longitudinally polarized, the axial density is nonzero. We can then define a meaningful 2D transverse mean-square axial radius in a generic EF for a longitudinally polarized target $s'=s$: 
\begin{equation}\label{EFaxialradius}
	\langle b_{A}^2 \rangle_\text{EF}(P_z) = \frac{\int\ud^2b_\perp\, b^2_\perp \,\mathcal A^0_\text{EF}(\uvec b_\perp;P_z)}{\int\ud^2b_\perp\,\mathcal A^0_\text{EF}(\uvec b_\perp;P_z)} = \frac{2}{3}\,R_A^2+ \frac{1}{2E_P^2}\YC{,}
\end{equation}
with $E_P=\sqrt{M^2+P_z^2}$. We stress, however, that the 2D axial density is not intrinsic and arises simply because a longitudinal boost mixes the components $\mathcal A^0$ and $\mathcal A_z$. In the LF formalism the boost is not necessary since one defines from the very beginning the density via the component $\mathcal A^+=(\mathcal A^0+\mathcal A_z)/\sqrt{2}$.

\begin{figure}[t!]
	\centering
	{\includegraphics[angle=0,scale=0.452]{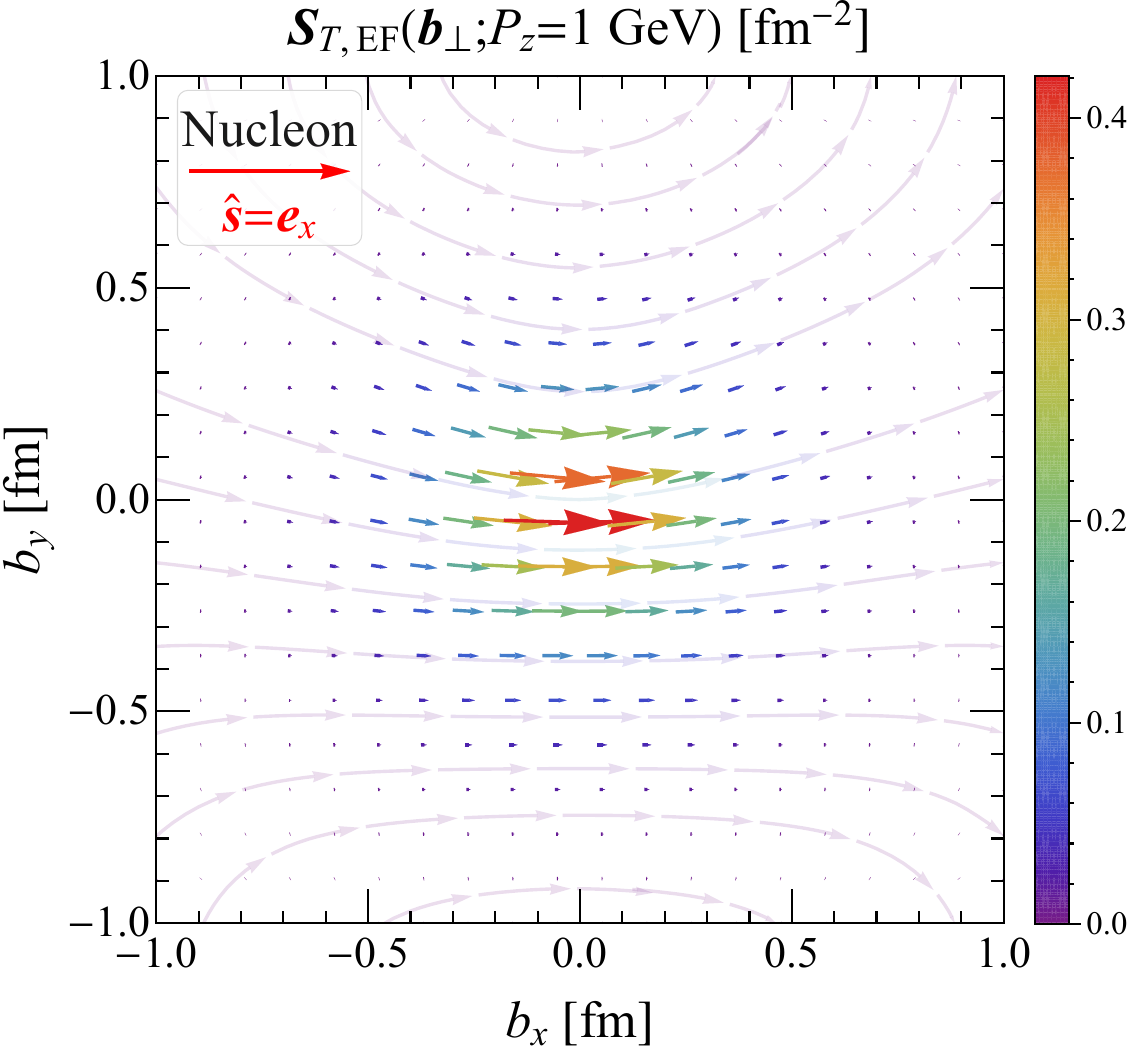}}
	\caption{The 2D transverse EF spin distribution $\uvec S_{T,\text{EF}}(\uvec b_\perp;P_z )=\frac{1}{2} \uvec{\mathcal A}_{\perp,\text{EF}}(\uvec b_\perp;P_z)$ from Eq.~\eqref{EF-def} for a transversely polarized nucleon at $P_z=1~\text{GeV}$, using the nucleon axial-vector FFs~\eqref{FFparamdip} and parameters from Table~\ref{Tab_FFs_Parameters}.}
	\label{Fig_2DEFSvT}
	\vspace{-3mm}
\end{figure}

Alternatively, we can consider the longitudinal and transverse EF spin distributions $S_{L,\text{EF}}\equiv \frac{1}{4} \text{Tr}[\sigma_z\, \mathcal A_{z,\text{EF}}] $ and $S_{T,\text{EF}}\equiv \frac{1}{4} \text{Tr}[\uvec\sigma_\perp \cdot \uvec{\mathcal A}_{\perp,\text{EF}}]= \hat{\uvec s}_\perp\cdot\uvec S_{T,\text{EF}}$, where the numerical result of $\uvec S_{T,\text{EF}}(\uvec b_\perp;P_z )$ is illustrated in Fig.~\ref{Fig_2DEFSvT}. The corresponding 2D transverse mean-square radii read
\begin{equation}
	\begin{aligned}\label{EF-spinradius}
        \langle b_{\text{spin},L}^2 \rangle_\text{EF}
        &= \frac{2}{3}\,R_A^2,\\
	\langle b_{\text{spin},T}^2 \rangle_\text{EF}(P_z) 
        &= \frac{2}{3}\,R_A^2 + \frac{1}{2 M^2}\frac{ G_P(0)}{G_A(0)}\\
        &\quad-\frac{1}{2M(E_P+M)}+\frac{1}{2 E_P^2}.
	\end{aligned}
\end{equation}
In the infinite-momentum limit $P_z\to\infty$, we recover, as expected, the LF transverse mean-square spin radii~\eqref{2DLF-SpinMSR-L} and~\eqref{2DLF-SpinMSR-T}. Interestingly, if we average over all spin directions and take $P_z\to 0$, we find
\begin{equation}
    \frac{\langle b_{\text{spin},L}^2 \rangle_\text{EF}+2\,\langle b_{\text{spin},T}^2 \rangle_\text{EF}(0) }{3}=\frac{2}{3}\,\langle r^2_\text{spin}\rangle,
\end{equation}
in agreement with the 3D BF mean-square spin radius given in Eq.~\eqref{def-new-spinMSR}, taking into account the fact that the latter does not depend on the spin direction. 
\newline

%
\textit{Conclusions.} We investigated in detail in this letter whether the widespread interpretation of $R_A^2$ given by Eq.~\eqref{def-Sachs-radius-Axial} as the nucleon mean-square axial radius is justified or not. In the Breit frame, where 3D spatial distributions are traditionally defined, the axial density vanishes identically~\eqref{3DBF-ampl}. This means that the 3D mean-square nucleon axial radius does not exist, and thus the interpretation of $R_A$ as the nucleon axial radius does not rest on sound physical grounds. In contrast, we demonstrated that a physically meaningful 3D nucleon mean-square spin radius~\eqref{def-new-spinMSR} can be unambiguously defined in terms of the same axial-vector FFs. 

Another option is to consider 2D spatial distributions in the light-front formalism. For a longitudinally polarized nucleon, the 2D transverse mean-square axial radius is equal to the transverse mean-square longitudinal spin radius, and is given by $\frac{2}{3}R^2_A$~\eqref{2DLF-SpinMSR-L}. For a transversely polarized nucleon, however, the transverse mean-square axial radius vanishes while the transverse mean-square transverse spin radius receives an additional contribution on top of the $\frac{2}{3}R^2_A$ term~\eqref{2DLF-SpinMSR-T}.

An interpolation between the Breit frame and light-front results can be obtained by using the concept of 2D elastic frame distributions, based on the phase-space formalism. When the nucleon has finite average momentum $P_z$, the 2D transverse mean-square axial and transverse spin radii receive additional $P_z$-dependent contributions compared to the light-front ones, see Eqs.~\eqref{EFaxialradius} and~\eqref{EF-spinradius}. These extra contributions formally disappear, however, in the infinite-momentum limit $P_z\to\infty$.
\newline

In summary, contrary to what is usually stated in the literature we did not find any proper justification for interpreting the quantity $R^2_A$ as the physical 3D mean-square axial radius of a spin-$\frac{1}{2}$ target (a nucleon for instance). On the other hand, the 3D mean-square spin radius $\langle r_\text{spin}^2 \rangle$~\eqref{def-new-spinMSR} is a well-defined and physically meaningful radius that characterizes more realistically the spatial extension of the weak content of the system. From a 2D perspective, however, the quantity $\frac{2}{3}R^2_A$ can consistently be interpreted as the 2D transverse mean-square axial or longitudinal spin radius, albeit only for a longitudinally polarized spin-$\frac{1}{2}$ target. 

While our analysis does not directly impact the phenomenology of axial-vector form factors, it explicitly demonstrates that the quantity $R^2_A$ \emph{alone} does not faithfully measure the spatial extension of the weak content of the nucleon. According to Eq.~\eqref{def-new-spinMSR}, the 3D mean-square spin radius $\langle r_\text{spin}^2 \rangle$ requires besides $R_A^2$ an additional determination of the ratio $G_P(0)/G_A(0)$, which provides an additional key motivation for ongoing lattice QCD calculations and future experimental measurements in (anti)neutrino-nucleon scatterings of the induced pseudoscalar form factor $G_P(Q^2)$.
\newline

%
\textit{Acknowledgments.} We warmly thank Raza~Sabbir~Sufian for very helpful communications at an early stage of this work, and Dao-Neng Gao and Ren-You Zhang for valuable discussions. This work is supported in part by the National Natural Science Foundation of China (NSFC) under Grant No.~12135011, by the Strategic Priority Research Program of the Chinese Academy of Sciences (CAS) under Grant No.~XDB34030102, and by the Chinese Academy of Sciences (CAS) under Grant No.~YSBR-101.

%
\bibliography{main_Refs.bib}
%
\end{document}